\documentclass{article}
\usepackage{graphicx}
\graphicspath{{converted_graphics/}}
\begin{document}

\begin{center}
{\LARGE Aerodynamical Effects in Snow Crystal Growth}\vskip12pt

{\large K. G. Libbrecht}\footnote{e-mail address: kgl@caltech.edu}\vskip4pt

{\large Department of Physics, California Institute of Technology}\vskip-1pt

{\large Pasadena, California 91125}\vskip-1pt

\vskip18pt

\hrule\vskip1pt \hrule\vskip14pt
\end{center}

\medskip \noindent \textbf{Abstract}: We review several aspects of
aerodynamics that affect the growth, morphology, and symmetry of snow
crystals. We derive quantitative estimates for aerodynamical forces that
orient falling snow crystals, estimate how air flow around snow crystals
affects their growth rates (the ventilation effect), and examine how the
combination of orientation and growth modification can stabilize or
destabilize different growth behaviors. Special attention is given to the
formation of triangular snow crystals, since it appears that aerodynamical
effects are responsible for producing this unusual morphology, both in
nature and in the laboratory.

\section{Introduction}

Complex patterns and structures often emerge spontaneously when crystals
grow, yielding a great variety of faceted, branched, and other forms. This
is readily seen, for example, in the well-known morphological diversity
found in naturally occurring mineral crystals \cite{minerals}. Suppressing
structure formation is often desired when growing large commercial crystals,
but exploiting the phenomenon provides a possible route for using nanoscale
self-assembly as a manufacturing tool \cite{nano}. Whether the goal is to
reduce, enhance, control, or simply understand structure formation, there
has been considerable interest from a number of fronts in characterizing the
detailed physical mechanisms that produce ordered structures from disordered
precursors during solidification \cite{solidification, shapes}.

The molecular dynamics involved in the condensation of disordered molecules
into a regular crystalline lattice is remarkably complex, involving a number
of many-body effects over different length scales and time scales \cite%
{saito}. As a result, calculating dynamical properties like crystal growth
rates from first principles is generally not yet possible. Predicting growth
morphologies has also proven quite difficult, with few overarching theories
connecting the many disparate physical mechanisms that govern growth
behaviors. A principal challenge is developing the theoretical tools that
connect detailed two-body molecular interactions to many-body crystal growth
dynamics.

A well-known and oft-studied example of structure formation during crystal
growth is the formation of ice crystals from water vapor. In the atmosphere
these are called snow crystals, and they fall from the clouds with a
remarkable diversity of morphologies, including simple plate-like and
columnar forms, elaborately branched plates, hollow columns, capped columns,
and many others \cite{fieldguide}. Surprisingly, a number of easily observed
aspects of snow crystal growth -- for example, that the growth morphology
depends strongly on temperature -- have not yet been adequately explained at
even a basic qualitative level (for a review, see \cite{libbrechtreview}).
We have been studying the detailed physics of snow crystals as a case study
in crystal growth, with the hope that developing a comprehensive mechanistic
model for this remarkably rich physical system will shed light on the more
general problem of structure formation during solidification \cite%
{libbrechtreview}.

The growth and morphology of snow crystals are mainly determined by two
processes -- the diffusion of water molecules through the surrounding air,
and the surface molecular dynamics (also called attachment kinetics) that
describes how water molecules join the crystalline lattice. Diffusion leads
to a growth instability that produces dendritic branching and other complex
structures, while attachment kinetics promotes the growth of smooth, faceted
structures. The two effects in concert yield the observed rich variety of
faceted and branched snow crystal morphologies. The physics of particle
diffusion is well understood, and numerical methods have been developed to
solve the diffusion equation and model the morphologies of growing crystals 
\cite{libbrechtmodel, gg, gg3d}. What is missing is an accurate
parameterization of the attachment kinetics that goes into these models, and
without this one cannot explain many aspects of the growth of snow crystals 
\cite{libbrechtreview}.

While diffusion and attachment kinetics are the dominant physical mechanisms
that describe snow crystal growth, several other processes can be important
as well, including: 1) the diffusion of latent heat generated at the growing
ice surface, 2) nucleation dynamics that may lead to twinned or other
polycrystalline structures, 3) chemical impurities on the ice surface that
affect the attachment kinetics, and 4) air flow around a growing crystal
that affect particle diffusion and thus alter growth rates. To obtain a
complete picture of snow crystal growth, and to accurately interpret
experimental observations, we must examine all the relevant physical
processes involved.

The present paper focuses on how aerodynamics affects snow crystal growth
and morphologies. This subject has been examined by a number of researchers
in the past (e.g. \cite{pruppacher, hallett, fukuta99, wang}), and we
examine four main areas here:

\noindent \textbf{Crystal Orientation} - how aerodynamic forces orient
falling snow crystals. We present derivations of quantitative estimates for
the various forces involved and estimate when air turbulence prevents snow
crystal orientation.

\noindent \textbf{The Ventilation Effect }- how crystal growth rates are
altered by air flow. From the theory for diffusion-limited growth, we derive
quantitative estimates for changes in growth as a function of air flow, and
we examine under what circumstances aerodynamics is an important factor.

\noindent \textbf{Snow Crystal Symmetry} - how aerodynamics can stabilize
growth rates to promote the growth of symmetrical snow crystals. We look at
how aerodynamics affects the formation of flat snow crystal plates, stellar
snow crystals with arms of equal length, and other morphologies.

\noindent \textbf{Triangular Snow Crystals} - how aerodynamics can produce a
growth instability that promotes the formation of triangular snow crystals.
We describe a model in which aerodynamical effects result in the formation
of snow crystals with trigonal morphologies that are readily seen in the
laboratory and in nature.

Our goal with this paper is to review the most relevant theoretical tools
and empirical results that describe how aerodynamics affects snow crystal
growth. While analytical and computational methods have been developed that
model fluid flow in great detail, we focus here on simple derivations and
rough approximations that give one an overarching physical understanding for
which processes are most important in different circumstances.

\section{Crystal Orientation}

The orientation of objects moving through fluids has been extensively
studied for many decades, and here we outline some basic results most
relevant to the case of snow crystals growing in air.

\subsection{Aerodynamic Drag}

At low Reynolds number, when there is no turbulence in the flow, we have 
\textit{Stokes drag} \cite{wikidrag}%
\begin{eqnarray*}
F_{Stokes} &=&bu \\
&=&6\pi \mu R_{H}u
\end{eqnarray*}%
where $u$ is the flow velocity, $\mu $ is the dynamic viscosity of the air,
and $R_{H}$ is the \textit{hydrodynamic radius} of the object. For a
spherical object, $R_{H}$ is equal to the radius $R$ of the sphere. (See
Appendix I for values of various physical quantities.)

At high Reynolds number, the fluid flow becomes turbulent and the drag is
proportional to the square of the velocity \cite{wikidrag}%
\[
F_{turb}=\frac{1}{2}C_{d}\rho _{air}Au^{2} 
\]%
where $A$ is the projected area of the object, $\rho _{air}$ is the air
density, and $C_{d}$ is a dimensionless drag coefficient. There is
considerable variation in $C_{d};$ a thin disk has a high value of $%
C_{d}\approx 1.15,$ while a streamlined body (essentially the cross-section
of a fish) has $C_{d}\approx 0.04$ \cite{dragcoeff}.

Note that we can derive the formula for turbulent drag in a crude fashion
rather simply from Newton's law. If we assume that air striking our object
essentially comes to a halt, then%
\begin{eqnarray*}
F &=&\frac{dp}{dt} \\
&\approx &\rho _{air}Au^{2}
\end{eqnarray*}%
which equals the above expression up to the numerical factor $C_{d}/2$.

At intermediate velocities one can use the formula for $F_{turb}$ with the
assumption that $C_{d}$ depends on the Reynolds number $R_{e},$ and detailed
parameterizations for $C_{d}(R_{e})$ are available \cite{wang}. For a rough
approximation we can take the total drag force as the sum of the two
components above, and for a disk this becomes (assuming $C_{d}\approx 1)$%
\begin{eqnarray*}
F_{drag} &\approx &F_{Stokes}+F_{turb} \\
&\approx &6\pi \mu Ru+\frac{1}{2}\rho _{air}\pi R^{2}u^{2}
\end{eqnarray*}%
where $R$ is the disk radius. The two components of $F_{drag}$ are equal
when $R_{e}=24,$ where we take

\begin{eqnarray}
R_{e} &=&\frac{2\rho _{air}uR}{\mu }=\frac{2uR}{\nu _{kin}}  \label{reynolds}
\\
&\approx &1.4\left( \frac{v}{10\textrm{ cm/sec}}\right) \left( \frac{R}{100%
\textrm{ }\mu \textrm{m}}\right)  \nonumber
\end{eqnarray}

\subsection{Terminal velocity}

The terminal velocity comes from equating%
\[
F_{drag}=mg
\]%
where $m$ is the mass of a falling snow crystal and $g$ is the gravitational
acceleration. (We assume $\rho _{ice}\gg \rho _{air}$.) Using both
components of $F_{drag}$ yields a quadratic equation for $v_{turb}.$ For
small crystals we can ignore $F_{turb}$ and the terminal velocity of a thin
disk of thickness $T$ becomes%
\begin{eqnarray*}
u_{term} &\approx &\frac{1}{6}\frac{\rho _{ice}g}{\mu }RT\textrm{ (thin disk;
low Reynold number) } \\
&\approx &8\left( \frac{R}{100\textrm{ }\mu \textrm{m}}\right) \left( \frac{T}{%
\textrm{10 }\mu \textrm{m}}\right) \textrm{ cm/sec }
\end{eqnarray*}%
For larger crystals we can ignore $F_{Stokes}$, giving%
\begin{eqnarray}
u_{term} &\approx &\sqrt{\frac{2T\rho _{ice}g}{\rho _{air}}}\textrm{ (thin
disk; high Reynold number) }  \label{fall1} \\
&\approx &40\left( \frac{T}{\textrm{10 }\mu \textrm{m}}\right) ^{1/2}\textrm{
cm/sec}  \nonumber
\end{eqnarray}%
The transition from Stokes to turbulent drag produces a shift from $%
v_{term}\sim RT$ to $v_{term}\sim T^{1/2},$ which occurs near%
\begin{eqnarray*}
R_{transition}^{2} &\approx &\frac{72\mu ^{2}}{Tg\rho _{ice}\rho _{air}}%
\textrm{ (thin disk)} \\
R_{transition} &\approx &450\left( \frac{\textrm{10 }\mu \textrm{m}}{T}\right)
^{1/2}\textrm{ }\mu \textrm{m}
\end{eqnarray*}%
In the high-Reynolds-number regime, Nakaya (\cite{nakaya}, page 114) found
fall velocities of close 30 cm/sec for plane dendritic crystals, independent
of $R$ for diameters ranging from 1.5 to 5 mm. Nakaya also observed (\cite%
{nakaya}, page 107) that these crystals had typical thicknesses of 9-15 $\mu 
$m, although disks with the same radius and mass would have effective
thicknesses of order 5-6 $\mu $m (\cite{nakaya}, page 115). These data agree
reasonably well with the expressions above, given the uncertainty in crystal
thicknesses.

For the case of a falling spherical snow crystal, the low Reynolds number
case gives%
\begin{eqnarray*}
u_{term} &=&\frac{2}{9}\frac{\rho _{ice}g}{\mu }R^{2}\textrm{ (sphere; low
Reynolds number)} \\
&\approx &1.1\left( \frac{R}{100\textrm{ }\mu \textrm{m}}\right) ^{2}\textrm{ m/sec%
}
\end{eqnarray*}%
while at high Reynolds number we have%
\begin{eqnarray*}
u_{term} &=&\sqrt{\frac{8}{3}\frac{R\rho _{ice}g}{\rho _{air}}}\textrm{
(sphere; high Reynolds number)} \\
&\approx &1.4\left( \frac{R}{100\textrm{ }\mu \textrm{m}}\right) ^{1/2}\textrm{
m/sec}
\end{eqnarray*}%
and the transition between these two behaviors occurs near%
\begin{eqnarray*}
R_{transition}^{3} &\approx &54\frac{\mu ^{2}}{g\rho _{ice}\rho _{air}}\textrm{
(sphere)} \\
R_{transition} &\approx &110\textrm{ }\mu \textrm{m}
\end{eqnarray*}

\subsection{Horizontal Alignment}

Aerodynamic forces often align falling snow crystals relative to the
horizontal plane; that is, plates tend to fall with their faces parallel to
the ground and columns fall with their long axes parallel to the ground. We
examine the case of thin plates here; the columnar case would be handled
similarly.

When the Reynolds number exceeds approximately $R_{e}\approx 100$, the
horizontal alignment of a falling plate is no longer stable, and a variety
of fluttering and tumbling modes occur \cite{naturedisks} . Using Equation %
\ref{fall1} for the terminal velocity, the criterion $R_{e}< 100$
gives 
\begin{eqnarray*}
R &<& \frac{R_{e,\max }\mu }{\sqrt{8T\rho _{air}\rho _{ice}g}} \\
R &<& 2\left( \frac{10\textrm{ }\mu \textrm{m}}{T}\right) ^{1/2}\textrm{ mm}
\end{eqnarray*}%
so we see that aerodynamic fluttering and tumbling instabilities are likely
to occur only for the largest snow crystals. Before these full-blown
instabilities develop, however, rocking or twirling motions are often
observed \cite{naturedisks} that produce deviations from horizontal that can
be substantial and time-dependent.

For small crystals with low Reynolds numbers, the limit to horizontal
alignment comes mainly from motion of the air, especially atmospheric
turbulence. (Brownian motion is likely less important than turbulence except
in the most stable laboratory conditions.) One effect of moving air is that
small snow crystals are simply carried along with the air. If a crystal
moves exactly with the air, however, then there is no relative motion of the
air with respect to the crystal, and thus no force that aligns the crystal.
Thus a perfectly steady wind will move a crystal, but will not change its
alignment. If air motion is going to affect alignment, then there must be a
relative velocity $u_{rel}$ between the air and the crystal. In Appendix II
we examine two models of atmospheric turbulence and use them to calculate
properties of $u_{rel}.$

The results from our turbulence models for thin plate crystals are shown in
Figure 1. Here we plot the terminal velocity $u_{term}$ along with $%
u_{turb,RMS},$ the root-mean-square value of $u_{rel}$ arising from air
turbulence. When $u_{term}$ becomes smaller than $u_{turb,RMS}$, then
aerodynamic forces will tend to align the crystals relative to the turbulent
air motion, and they will no longer be aligned with the horizontal plane.

When falling snow crystals have simple, faceted morphologies and are aligned
to the horizontal plane, then conditions are favorable for the formation of
light pillars and some types of atmospheric halos \cite{lightcolor}. Halo
observers have found that crystals smaller than $R\approx 25$ $\mu $m show
little horizontal alignment in the atmosphere \cite{tape1, tape2}, and this
agrees with the results shown in Figure 1 (albeit with considerable
uncertainty in the turbulence models). Remarkably, crystal alignments within
one degree of horizontal have been inferred from halo observations \cite%
{tape1}, but only in very rare circumstances.

\begin{figure}[ht] 
  \centering
  \includegraphics[width=4.5in,keepaspectratio]{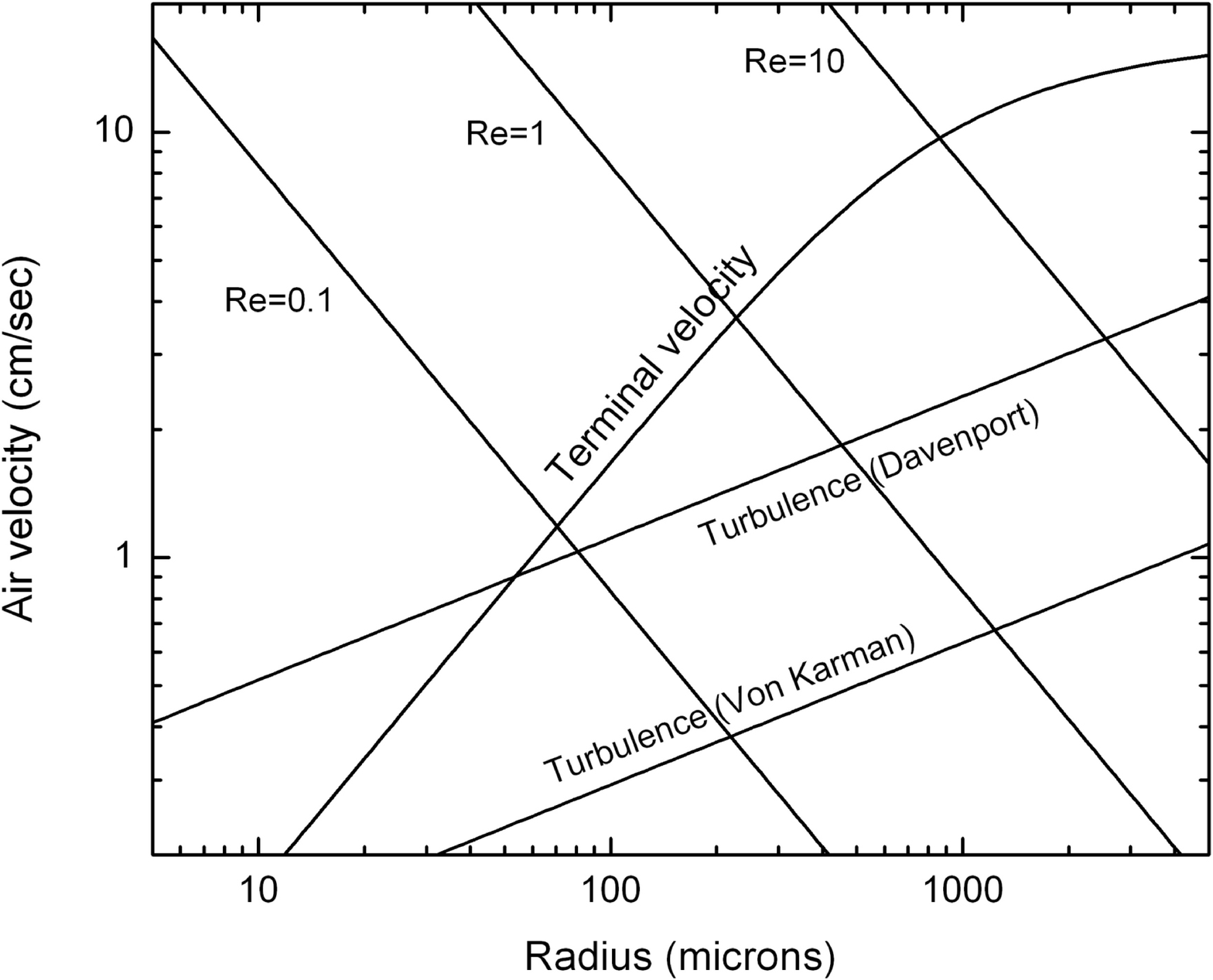}
  \caption{The velocity of a 2-$\protect%
\mu $m-thick snow crystal disk relative to the ambient air surrounding the
crystal, as a function of the radius of the disk. The \textit{terminal
velocity} curve shows the fall velocity of the crystal in still air. The 
\textit{turbulence} curves show the RMS velocity arising from air turbulence
when the mean air speed is 1 m/sec, from Equation \protect\ref{turb1}, for
the two turbulence models described in Appendix II. Other lines show
different Reynolds numbers from Equation \protect\ref{reynolds}. When $R$ is
large enough that the terminal velocity curve is above the turbulence
curves, then gravity will align the crystal horizontally. Otherwise
turbulent air motion will disturb the horizontal alignment. The graph for
thicker disks looks similar, but the terminal velocity is higher relative to
that from turbulence.}
  \label{velocities}
\end{figure}

\section{The Ventilation Effect}

When there is air flowing around a snow crystal, its growth rate increases
as the flow enhances particle and heat diffusion in the neighborhood of the
crystal. The increase in crystal growth is often called the \textit{%
ventilation effect} (for a review, see \cite{pruppacher}). In this paper we
will ignore heat diffusion, since it is typically less important than
particle diffusion in limiting snow crystal growth \cite{libbrechtreview}.
Removing latent heat generated by solidification becomes a relatively more
important factor in laboratory experiments at lower pressures, but here we
are mainly concerned with aerodynamic effects at pressures one might
encounter under normal atmospheric conditions.

To examine how air flow affects particle diffusion, consider that the
timescale for water molecules to diffuse a distance $L$ through the air is%
\[
\tau _{diffusion}\approx \frac{L^{2}}{D} 
\]%
where $D$ is the diffusion constant (see Appendix I). A growing crystal
significantly reduces the supersaturation in its vicinity only out to
distances comparable to the overall crystal size \cite{libbrechtreview}, so
we typically take $L$ in the above expression to be the approximate size of
the crystal. The timescale for air to flow past an object of size $L$ is 
\[
\tau _{flow}\approx L/u 
\]%
where $u$ is the velocity of the flow. For a snow crystal growing in air,
and when the growth is limited by particle diffusion, we would expect the
flow to significantly affect the growth only when $\tau _{flow}<\tau
_{diffusion},$ or equivalently when $u>D/L$ or $R_{e}>1$ (where we have used 
$D\approx \nu _{kin}$; see Appendix I).

\subsection{The Spherical Case}

To a rough approximation, the effect of a slow relative velocity between a
growing crystal and the surrounding air can be modeled by moving the
far-away boundary condition from infinity to some finite outer boundary $%
R_{outer},$ at which the supersaturation is $\sigma _{outer}.$ With no air
motion we revert to $R_{outer}\rightarrow \infty $ and $\sigma
_{outer}\rightarrow \sigma _{\infty }.$ This is clearly a crude model, since
it does not include any asymmetries resulting from the air flow, but it is
useful for determining the magnitude of the ventilation effect under
different conditions. Assuming this change in effective boundary, we can
solve the spherically symmetric case exactly. The supersaturation becomes 
\[
\sigma \left( r\right) =A+\frac{B}{r} 
\]%
which satisfies the diffusion equation, and the outer boundary condition
gives%
\[
\sigma _{outer}=A+\frac{B}{R_{outer}} 
\]%
while inner boundary condition becomes%
\begin{eqnarray*}
v &=&\frac{c_{sat}D}{c_{ice}}\frac{d\sigma }{dr}(R)=-\frac{c_{sat}D}{c_{ice}}%
\frac{B}{R^{2}} \\
&=&\alpha v_{kin}\sigma \left( R\right) =\alpha v_{kin}\left( A+\frac{B}{R}%
\right)
\end{eqnarray*}%
where $v$ is the crystal growth velocity, $\alpha $ is the attachment
coefficient, $c_{sat}$ is the water vapor density in saturated air, $c_{ice}$
is the solid density, and $v_{kin}$ is a constant term \cite{libbrechtreview}%
. Solving these equations for $A$ and $B$ yields 
\[
v=\frac{\alpha \alpha _{diff}}{\alpha \beta +\alpha _{diff}}v_{kin}\sigma
_{outer} 
\]%
where%
\[
\alpha _{diff}=\frac{c_{sat}D}{c_{ice}v_{kin}R} 
\]%
and%
\[
\beta =\frac{R_{outer}-R}{R_{outer}}<1 
\]

If $R_{outer}\rightarrow \infty ,$ then we can take $\beta =1$ and this
expression reverts to the usual result for a growing spherical crystal \cite%
{libbrechtreview}. If $\alpha \ll \alpha _{diff}$ then the growth is mainly
limited by attachment kinetics and we have%
\[
v\approx \alpha v_{kin}\sigma _{outer} 
\]%
In this case the growth is independent of $\beta ,$ as we would expect. If $%
\alpha _{diff}\ll \alpha ,$ so the growth is primarily diffusion limited,
then we have%
\[
v\approx \beta ^{-1}\alpha _{diff}v_{kin}\sigma _{outer}\approx \beta ^{-1}%
\frac{c_{sat}D}{c_{ice}R}\sigma _{outer} 
\]%
which is larger than the usual result for diffusion-limited growth by a
factor of $\beta ^{-1}$.

From the discussion of timescales above, the change in $R_{out}$ caused by
air flow can be approximated by taking%
\[
R_{out}\approx R+\frac{D}{u}
\]%
for the spherical case, which gives the growth enhancement factor 
\begin{equation}
f_{v}=\beta ^{-1}\approx 1+\frac{uR}{D}\approx 1+aR_{e}  \label{enhancement}
\end{equation}%
where $a$ is a dimensionless geometrical factor. Thus the growth rate is
increased by a factor of $f_{v}=\left( 1+aR_{e}\right) $ when the flow
velocity is small and the growth is mainly diffusion-limited. Detailed
modeling and experiments with spherical water drops suggests $a\approx 0.1$
when $R_{e}<1$ \cite{pruppacher}.

For higher $R_{e},$ the enhancement factor contains a term proportional to $%
R_{e}^{1/2}$ instead of $R_{e},$ again based on numerous calculations and
experiments relating to the growth and evaporation of spherical water
droplets \cite{pruppacher}. Thus we have a rough approximation for the
growth enhancement factor%
\begin{eqnarray*}
f_{v} &\approx &1+0.1R_{e}\textrm{ (for }R_{e}<1) \\
f_{v} &\approx &0.8+0.3R_{e}^{1/2}\textrm{ (for }R_{e}>1)
\end{eqnarray*}%
which applies to the spherical case when the growth is mainly diffusion
limited. This semi-empirical result is of limited use for snow crystal
growth, since snow crystals are generally not very spherical and their
growth is often substantially limited by attachment kinetics. 

\subsection{Dendrite Growth}

The growth of a snow crystal dendrite must be considered differently, since
there is no isolated far-away boundary as there is with the spherical case.
Assuming a simple parabolic shape (a reasonable first approximation to a
dendrite tip), with pure diffusion-limited growth, the diffusion equation
can be solved to give the tip growth velocity 
\[
v_{tip}\approx \frac{2D}{R_{tip}\log (\eta _{\infty }/R)}\frac{c_{sat}}{%
c_{solid}}\sigma _{\infty }
\]%
where here $R_{tip}$ is the radius of curvature of the dendrite tip and $%
\eta _{\infty }$ is the distance to the outer boundary in parabolic
coordinates \cite{cloudy, libbrechtreview}. If we assume the air flow is
directed into the dendrite tip along the symmetry axis, and thus take $\eta
_{\infty }\approx R+D/u$ following the discussion above, then 
\begin{eqnarray*}
v_{tip} &\approx &\frac{2D}{R_{tip}\log (1+1/R_{e,tip})}\frac{c_{sat}}{%
c_{solid}}\sigma _{\infty } \\
&\approx &\frac{-2D}{R_{tip}\log (R_{e,tip})}\frac{c_{sat}}{c_{solid}}\sigma
_{\infty }
\end{eqnarray*}%
where the latter expression assumes $R_{e,tip}\approx R_{tip}u/D\ll 1.$ As
an example, if we take $R_{tip}\approx 10$ $\mu $m for the dendrite tip
radius, then $R_{e,tip}$ increases from $\approx 0.015$ to $\approx 0.035$
when the air velocity increases from $u=3$ cm/sec to $u=7$ cm/sec, which
increases $-\log (R_{e,tip})^{-1}$ by roughly 25 percent. This is in
reasonable agreement with measurements (\cite{hallett}, Figure 8).

It is interesting to compare this result with the spherical case; that is, a
sphere with $R\approx 10$ $\mu $m growing in the presence of flows with $u=3$
cm/sec and $u=7$ cm/sec. In the spherical case the outer boundary goes from $%
R_{out}\approx 700$ $\mu $m to $R_{out}\approx 300$ $\mu $m, producing a
mere 2\% increase in growth rate, even assuming a rather large $\beta
^{-1}\approx 1+R_{e}$. For the dendrite case, changing $\eta _{\infty }$
from the same $700$ $\mu $m to $300$ $\mu $m produces a growth increase of
25\% -- more than an order of magnitude larger. Physically, this
demonstrates how much the main body of a parabolic dendrite affects growth
near the tip. Even a low-velocity flow directed at the dendrite tip
effectively removes a large part of the dendrite body from the diffusion
problem and thus substantially increases the tip growth rate.

\begin{figure}[t] 
  \centering
  \includegraphics[width=5.4in,keepaspectratio]{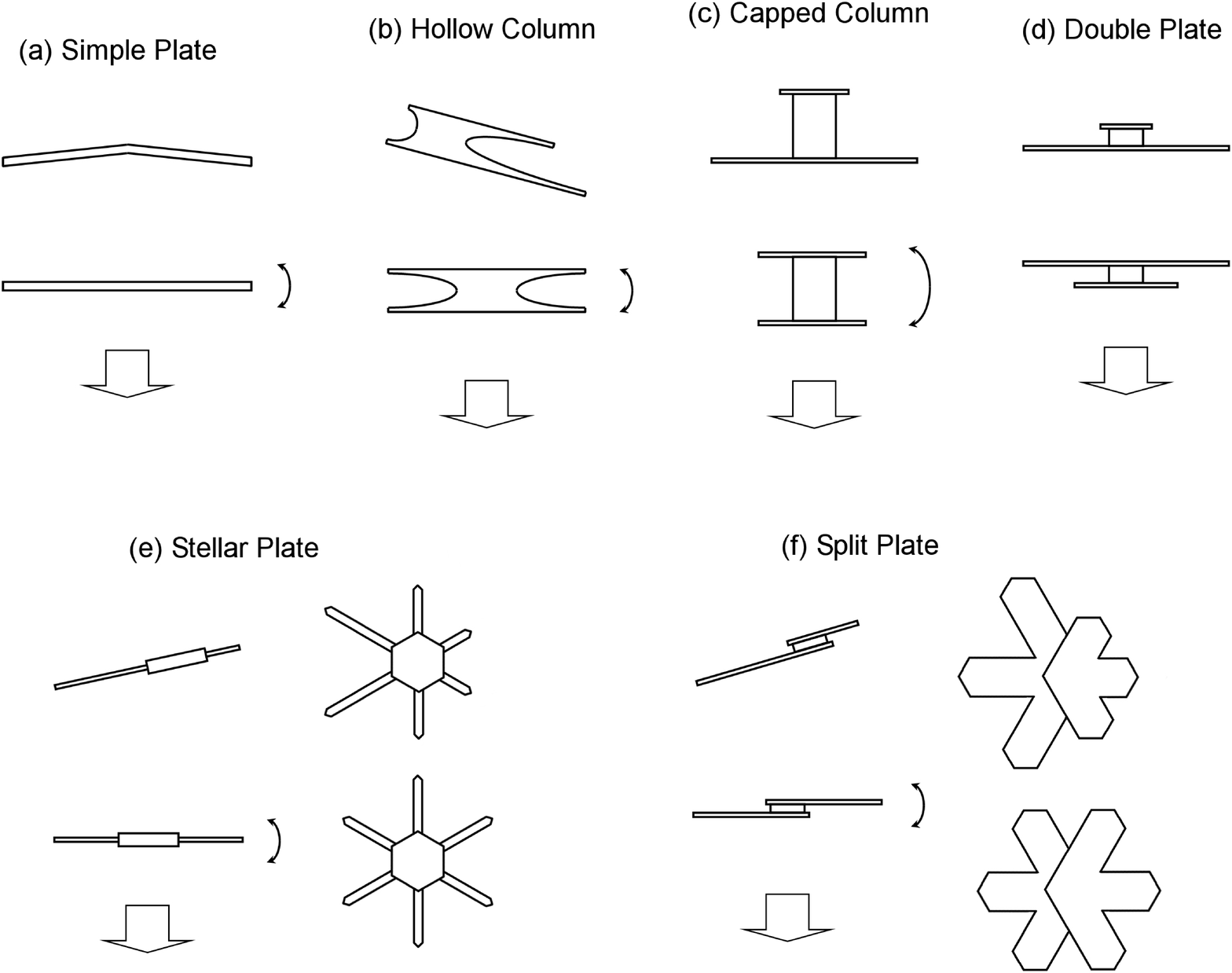}
  \caption{Schematic drawings of different
types of snow crystals, showing how aerodynamic effects often tend to
improve the overall crystal symmetry. For each case, the upper drawing shows
how the crystal might grow in a fictional "negative" world, in which
aerodynamic effects are opposite to those found in reality; the lower
drawing in each pair shows how the same crystals grow in the real world. The
large arrow shows the fall direction for the crystal; the curved arrows
indicate that the crystal flips over as it grows. More details are described
in the text.}
  \label{xtals}
\end{figure}

\section{Snow Crystal Symmetry}

Given the various calculations described above, we now examine how
aerodynamics can affect the symmetry of growing snow crystals. In many
instances the alignment of falling crystals combines with the ventilation
effect to balance the growth of different parts of a crystal, thus enhancing
its overall symmetry. Although this symmetry-enhancing effect is relatively
weak, it often has noticeable effects.

In Figure 2 we show several different snow crystal types (described in \cite%
{fieldguide}) that are influenced by aerodynamics. To illustrate the
discussion, this figure includes drawings that show how these crystals grow
in the real world along with drawings showing how they might grow in a
fictional \textquotedblleft negative\textquotedblright\ world, where the
aligning effects of aerodynamics have the opposite sign compared to the real
world. We consider each case in Figure 2:

\textbf{Simple Plates}. In Figure 2a, a falling hexagonal plate crystal is
aligned with respect to the horizontal plane, so the crystal sees a flow of
air directed at its lower face. This flow enhances diffusion via the
ventilation effect so the lower face grows more rapidly than the top face.
Since the outer parts of the basal faces grow more rapidly than the center
of the plate (because the edges stick out farther into the supersaturated
air), this causes the plate to grow with a slightly conical shape, as shown
in the top diagram in Figure 2a. In our fictional negative world,
aerodynamics keeps the plate aligned with its starting orientation, so it
continues growing in the shape of a shallow cone.

In the real world, however, the conical shape drawn in Figure 2a is
aerodynamically unstable and the crystal soon flips over, causing the point
of the nascent cone to face downward. Subsequent growth then tends to
reverse the initial conical shape. Aerodynamic alignment and ventilation
effect thus cause the crystal to flip numerous times as it grows, resulting
in a plate that is very close to being perfectly flat. This scenario of a
constantly flipping plate has been witnessed in observations of growing
crystals where an applied airflow was used to counter the downward drift
from gravity \cite{fukuta99}.

We see here that aerodynamic effects enhance the symmetry of the crystal --
in this case ensuring that the plate remains flat. The ventilation effect
first results in the preferential growth of the lower face of the plate,
thus breaking the symmetry, but this in turn produces an aerodynamic
instability that flips the crystal and restores symmetry.

\textbf{Hollow Columns}. If we consider the growth of a hollow column along
its length (see Figure 2b), we see that there is no initial symmetry
breaking growth that favors one end over the other. Nevertheless, random
perturbations will lead to small differences as the crystal grows. In our
negative world, the longer end would tip downward and the ventilation effect
would cause this end to grow faster in comparison to the opposite end. The
result would be a growth instability favoring asymmetrical growth. In the
real world, however, aerodynamic forces cause the initially longer end to
tip upward (because the center of the crystal is heavier than the hollow
ends). Then the ventilation effect favors the short end, and the net effect
is to restore symmetry between the two ends.

We see another effect with the growth of the sides of the column. As the
column falls with its axis horizontal, the ventilation effect causes the
lower parts of the crystal to grow more rapidly than the upper parts.
Eventually the crystal becomes sufficiently asymmetrical that it flips over,
thus reducing growth on the heavy side. Again with the radial growth of
columns, we see that aerodynamic forces and the ventilation effect conspire
to enhance the morphological symmetry of the crystal.

\textbf{Capped Columns}. A capped column may align (depending on its
different dimensions) with one end-plate facing downward, as shown in Figure
2c. In our negative world, the lower plate would grow more rapidly and would
soon dominate over the upper plate. In the real world, any asymmetry in
growth causes an aerodynamic instability that flips the column over. As with
the previous examples, multiple flips again promote the growth of a
symmetrical crystal. This behavior has also been seen experimentally \cite%
{fukuta99}.

\textbf{Double Plates}. If the two ends of a capped column are close
together, then a diffusion-induced instability may win out over aerodynamic
effects, causing one plate to dominate over the other. This time our
negative and real worlds do not produce such different results, as shown in
Figure 2d. The crystal falls with the dominant plate facing down in the
negative case, and the secondary plate is almost completely shielded from
additional growth. In the real world, the smaller plate faces downward, so
the ventilation effect favors its growth. The dominant plate still
overshadows its smaller companion from the standpoint of vapor diffusion, so
symmetry is not preserved, in contrast to the capped column case.

\textbf{Stellar Plates}. If the center of a stellar plate is more massive
(per unit area) than the outer parts, then aerodynamic forces will cause the
plate to tip if it grows asymmetrically. In our negative world, the longer
branches would tip downward and the ventilation effect would favor their
growth, resulting in a growth instability that increases the asymmetry of
the crystal.

In the real world, however, the long branches would tip upward, where their
growth would slow relative to the shorter branches. This would tend to
restore the symmetry of the crystal, as shown in Figure 2e, and the crystal
orientation would remain flat as it falls.

This symmetry-restoring effect is likely very weak, and it would be
especially weak in the case of a fern-like stellar dendrite, which typically
does not have a heavy core. In the absence of a heavier center, the crystal
would continue to orient close to horizontally even if some branches were
longer than others, as long as the overall shape is roughly disk-like.

It should be emphasized that in the absence of any of the aerodynamical
effects described here, the six arms of a stellar plate crystal would still
grow symmetrically. The standard model for the formation of stellar
crystals, in which the arms all grow nearly identically because they
experience essentially identical time-dependent conditions during their
growth \cite{libbrechtreview}, adequately explains the occasionally high
degree of six-fold symmetry observed in these crystals.

\textbf{Split Plates}. A split plate crystal is a short capped column where
one end-plate dominates on one side of the crystal and the other end-plate
dominates on the other side \cite{fieldguide}, as illustrated in Figure 2f.
The heavy center, along with the separation between the two plates, leads to
a relatively strong aerodynamic instability in these crystals. In our
negative world, the crystal would quickly become asymmetrical, as shown in
the Figure. In the real world, however, the crystal flips as soon as one
side grows appreciably larger than the other. As with the other examples,
repeated flips result in a crystal with good symmetry.

Close inspection of symmetrical stellar snow crystals reveals that a
surprising number are actually split plates \cite{fieldguide}. Aerodynamic
affects may well play a substantial role in enhancing the symmetry of these
initially asymmetrical crystals, thus explaining why split plates often look
so much like normal stellar plates.

\begin{figure}[ht] 
  \centering
  \includegraphics[width=4.2in,keepaspectratio]{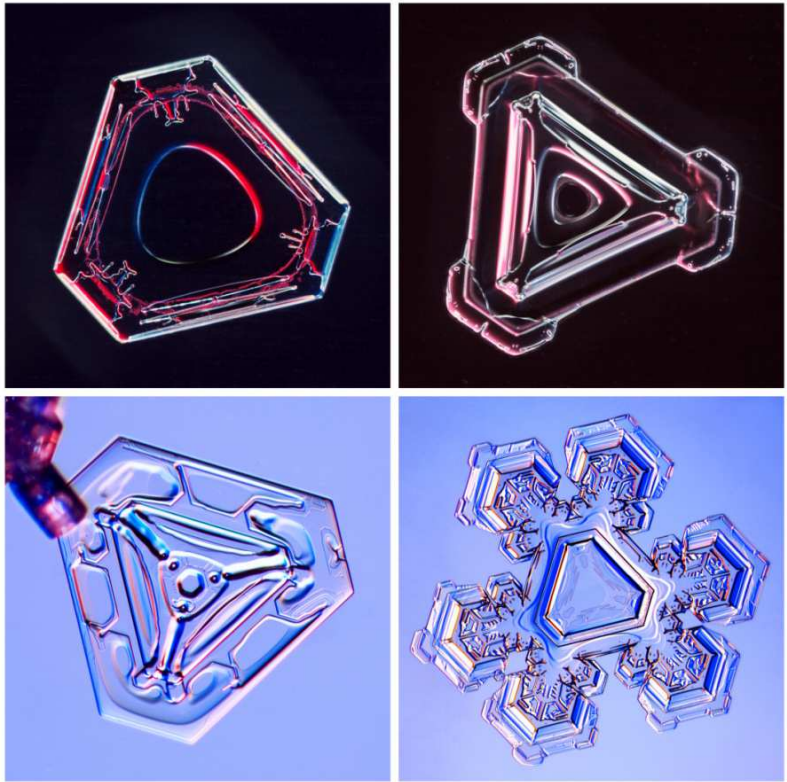}
  \caption{Examples of natural snow
crystals exhibiting triangular morphologies (from \protect\cite{snowflakes}%
). The equivalent diameters (defined by $d=(4A/\protect\pi )^{1/2}$, where $%
A $ is the projected 2D crystal area) range from 1 to 3 mm. The lower right
example shows a crystal with an initial truncated triangular morphology
(outlined by the central surface markings) that subsequently grew plate-like
branches.}
  \label{triangulars}
\end{figure}

\section{Triangular Snow Crystals}

An especially interesting application of aerodynamic effects in snow crystal
growth is the explanation of triangular snow crystals \cite{libbarnold}.
Figure \ref{triangulars} shows several examples that illustrate some typical
characteristics of this morphological type. The most common occurrence is a
small plate-like crystal that has the overall shape of a truncated
equilateral triangle, often with a variety of surface markings on the basal
faces \cite{fieldguide}. Some truncated triangular plates sprout branches as
they grow larger, as shown in the lower right example in Figure \ref%
{triangulars}.

Observers of natural snow crystals have reported the occurrence of
triangular snow crystals for nearly two centuries. The first documented
observation (to my knowledge) was by William Scoresby in 1820 \cite{scoresby}%
, and Wilson Bentley and W. J. Humphreys presented several dozen examples
with triangular morphologies in their well-known 1931 compilation of snow
crystal photographs \cite{bentley}. Although triangular crystals are usually
small and relatively rare, they are visually distinctive and fairly easy to
find in nature. Some especially large and cleanly faceted triangular
crystals have been recorded at the South Pole \cite{tape1, tape2}.

\begin{figure}[t] 
  \centering
  \includegraphics[width=3.2in, keepaspectratio]{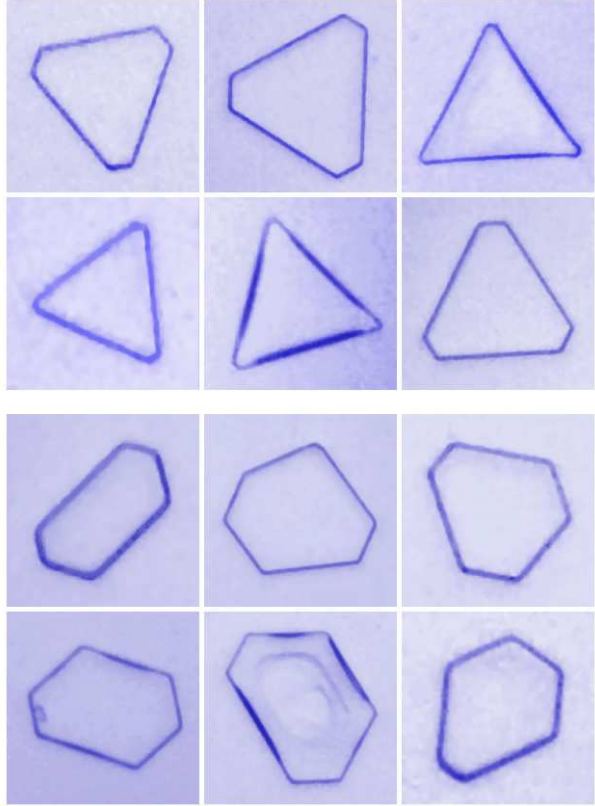}
  \caption{Examples of some extreme ($H
< 1/3$) snow crystal plates grown at -10 C with a
supersaturation of 1.4 percent and an air pressure of one bar. The top six
images show crystals with $T < 1/2$, while the bottom six show
crystals with $T > 1/2$. The former are generally much more
common than the latter. Equivalent diameters of these crystals range from
roughly 50 to 100 microns \protect\cite{libbarnold}.}
  \label{extremextals}
\end{figure}

\subsection{Laboratory-Grown Crystals}

\begin{figure}[t] 
  \centering
  \includegraphics[width=5.5in,keepaspectratio]{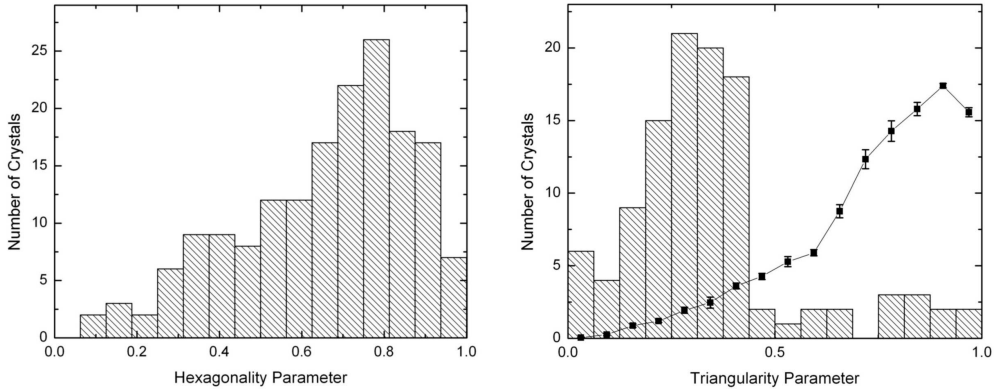}
  \caption{(Left) Distribution of
laboratory-grown crystals as a function of the hexagonality parameter $H$
defined in the text (from \protect\cite{libbarnold}). These data are from an
unbiased sample of plate-like crystals grown in free-fall in air at -10 C
with a water vapor supersaturation of 1.4 percent. (Right) Distribution of
laboratory grown crystals with $H<1/3$, as a function of the triangularity
parameter $T$ defined in the text. The line shows a Monte Carlo model for $T$
that assumes random growth perturbations of a hexagonal plate. Error bars
show the uncertainty in the model, estimated by varying a number of details
in the calculations. The data and model show that crystals with a triangular
morphology (small $T$) are much more common than one would expect from
random growth perturbations.}
  \label{tandh}
\end{figure}

Triangular morphologies have been observed occasionally in laboratory
studies of snow crystals that have been grown while falling in air (e.g. 
\cite{yamashita, libbarnold}), and some examples are shown in Figure \ref%
{extremextals}. To examine the distribution of different morphologies in a
systematic fashion, Libbrecht and Arnold \cite{libbarnold} defined a
\textquotedblleft hexagonality\textquotedblright\ parameter $H=L_{1}/L_{6}$
for faceted snow crystal plates, where $L_{1}$ and $L_{6}$ are the lengths
of the shortest and longest prism facets, respectively. $H$ is close to
unity when a plate is nearly hexagonal, while $H$ is smaller for any of a
variety of odd-shaped plates. Figure \ref{tandh} shows the measured $H$
distribution for an unbiased sample of their data. Crystals with $H>3/4$
appeared nearly hexagonal to the eye, and this plot confirms that simple
plate-like crystals are mostly hexagonal in shape.

Libbrecht and Arnold also defined a \textquotedblleft
triangularity\textquotedblright\ parameter $T=L_{3}/L_{4}$, where $L_{3}$
and $L_{4}$ are the lengths of the third and fourth smallest facets,
respectively. This parameter is small if, and only if, the morphology is
that of a truncated triangle, and $T\rightarrow 0$ if, and only if, the
morphology is nearly that of an equilateral triangle. Figure \ref{tandh}
shows the $T$ distribution using \textquotedblleft
extreme\textquotedblright\ crystals (defined as those with $H<1/3)$. That
the $T$ distribution is skewed to lower values reflects that fact that most
of the extreme crystals were shaped like truncated triangles.

Libbrecht and Arnold then used Monte Carlo simulations to investigate the
null hypothesis that their observations were due entirely to random
variations in the growth rates of the different prism facets, yielding the
line in Figure \ref{tandh}. In these simulations, random fluctuations in the
growth rates of the six facets produced a variety of odd shapes, including
trapezoidal, diamond-shaped, and other forms. However, as seen in Figure \ref%
{tandh}, the simulations did not show a preponderance of triangular crystals
over other non-hexagonal shapes, in stark contrast to the data.

These considerations provide convincing evidence that, at least under some
growth conditions, triangular morphologies are substantially more abundant
than one would expect from random fluctuations in the growth of hexagonal
plates. Once can then conclude that some non-random mechanism is responsible
for the growth of snow crystal plates with three-fold symmetry. In
particular, this mechanism must somehow coordinate the growth of the facets
so that they alternate between slow and fast growth around the crystal.

\subsection{An Aerodynamic Model}

\begin{figure}[t] 
  \centering
  \includegraphics[width=2.2in,keepaspectratio]{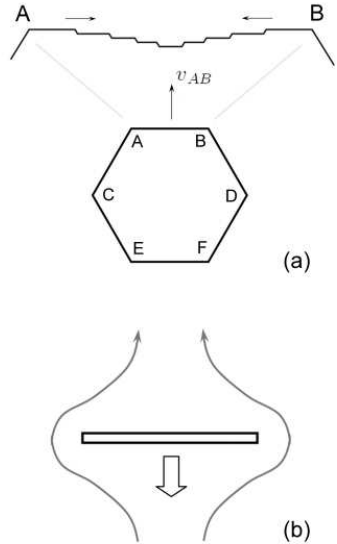}
  \caption{(a) The formation of facets in a
simple hexagonal plate-like crystal, as described in the text. The top
figure shows a close-up of the $AB$ facet, exaggerated to show molecular
steps on the surface. (b) A schematic depiction of airflow around a falling
hexagonal plate crystal (seen from the side).}
  \label{hex1}
\end{figure}

Libbrecht and Arnold proposed an aerodynamic model to explain the growth of
triangular snow crystals, both in the lab and in nature \cite{libbarnold},
which we examine here. To begin, first consider the case of a thin hexagonal
plate crystal, as shown in Figure \ref{hex1}. The crystal has six prism
facets, and each grows outward at some perpendicular growth velocity (in
Figure \ref{hex1}, for example, $v_{AB}$ is the growth velocity of the $AB$
facet). For a symmetrical crystal, all six prism facets are the same length
and all six growth velocities are equal.

The growth of a faceted crystal is limited partially by water vapor
diffusion through the surrounding air and partially by attachment kinetics
at the crystal surface. The two effects together result in facet surfaces
that are slightly concave at the molecular level, as shown in Figure \ref%
{hex1}, although they may appear perfectly flat optically \cite{saito}.
Nucleation of new molecular terraces occurs near the corners (points $A$ and 
$B$ in Figure \ref{hex1}), where the supersaturation is highest. The
molecular steps then propagate inward, traveling more slowly near the facet
centers where the supersaturation is lower. The combined effects of surface
attachment kinetics and diffusion-limited growth thus automatically
establish the concave shape of each facet surface.

The perpendicular growth velocity $v_{AB}$ of the $AB$ facet is primarily
determined by the nucleation rates near points $A$ and $B$. In the
symmetrical case, the rates at $A$ and $B$ are equal, giving the picture
shown in Figure \ref{hex1}. If the nucleation rate were slightly greater at $%
A$, then the picture would be distorted and the facet surface would be
tilted slightly relative to the ice lattice. If the nucleation rate at $A$
were substantially greater than at $B$, then terraces generated at $A$ would
propagate all the way across a vicinal surface from $A$ to $B$ \cite{saito}.
In general, we see that $v_{AB}$ is determined by the greater of the two
nucleation rates at $A$ and $B$.

As a growing crystal falls, air resistance causes the basal faces to become
oriented perpendicular to the fall velocity, as shown in Figure \ref{hex1}.
Air flowing around the falling crystal tends to increase growth via the
ventilation effect, and the airflow produces an effective increase in
supersaturation where the edges stick out farthest and the resulting flow is
the fastest. In the case of a falling plate, this aerodynamic effect mostly
increases the growth of the thin edges of the plate (i. e., the prism
facets).

\begin{figure}[t] 
  \centering
  \includegraphics[width=3in,keepaspectratio]{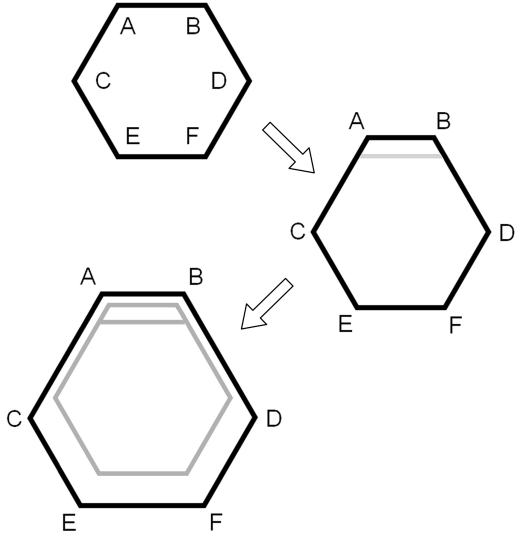}
  \caption{An aerodynamic model for the
formation of triangular crystals, as described in the text.}
  \label{hex2}
\end{figure}

With this overall picture in mind, we now consider the case shown in Figure %
\ref{hex2}. Here we start with a hexagonal plate and assume a small growth
perturbation somewhere on the $AB$ facet that breaks the six-fold symmetry
and makes $v_{AB}$ greater than the other five facet growth rates. This
perturbation could come from a crystal dislocation, a step-generating
chemical impurity on the surface, a piece of dust, or perhaps some other
mechanism. Regardless of the origin of the initial symmetry-breaking
perturbation, the larger growth rate $v_{AB}$ initially results in the
distorted crystal shape shown as the second stage in Figure \ref{hex2}.

Once this initial asymmetry appears, the additional material produces an
increased aerodynamic drag on one side of the plate. This in turn changes
the orientation of the falling crystal in such a way that the points $E$ and 
$F$ tip downward, toward the fall direction. Relative to the original
horizontal orientation, air flowing around the tilted crystal now increases
the supersaturation at points $E$ and $F$ while decreasing it at $A$ and $B$%
. As a result, the nucleation rates at points $E$ and $F$ increase. Air
turbulence will upset the horizontal orientation of the crystal, but it will
enhance this ventilation effect. Whatever the source of a relative velocity
between crystal and air, aerodynamic forces will align the crystal relative
to the flow.

The growth rate of a facet is determined mainly by the greater of the
nucleation rates at its two corners (from the discussion of Figure \ref{hex1}
above), and for our tilted crystal we see that the nucleation rates at $E$
and $F$ must be greater than at $C$ and $D$. Thus we must have that $%
v_{CE}\approx v_{EF}\approx v_{DF}$ to a rough approximation, and
furthermore these velocities must all be greater than $v_{AC}$ and $v_{BD}$
(while $v_{AB}$ is somewhat ill-determined throughout because of the assumed
growth perturbation). After a period of additional growth in these
conditions, shown in the last stage of Figure \ref{hex2}, we find that the
length of facet $EF$ has increased relative to $CE$ and $DF$. This means
that $AB$, $CE$, and $DF$ are now the three shortest facets, so the crystal
has begun to assume a slight triangular shape.

For small crystals like those grown in the laboratory, the ventilation
effect produces only a rather small change in growth rates of the different
facets. From Figure \ref{velocities} we see that the Reynolds number for
small laboratory-grown crystals is roughly $R_{e}\approx 0.1,$ so the
enhancement factor from Equation \ref{enhancement} gives at most about a 10
percent difference in growth rates for the different facets. This
enhancement factor alone is probably too small to explain the triangular
morphologies seen, but several other effects can further augment the
differences between the facets. First, once aerodynamics produces the shape
shown in the last stage of Figure \ref{hex2}, the shorter facets ($AB$, $CE$%
, and $DF$ in this case) stick out farther into the supersaturated air, so
the Mullins-Sekerka instability \cite{saito} tends to increase $v_{AB}$, $%
v_{CE}$, and $v_{DF}$ relative to the other three facets. An important
aspect of this instability is that it takes less mass to grow a surface of
reduced size, so overall mass flow considerations in diffusion-limited
growth tend to increase the growth of the shorter facets. Second, airflow is
faster around the shorter facets, because they stick out farther, and this
also increases their growth relative to the longer facets via the
ventilation effect. Third, if the condensation coefficient is a strong
function of supersaturation, as is often the case, then even rather small
changes in effective supersaturation from the ventilation effect could
result in substantial changes in relative growth rates.

The end result in this model is that an initial symmetry-breaking
perturbation results in a growth morphology that becomes more triangular
with time. Only one initial perturbation is necessary, and no coordination
intrinsic to the molecular structure of the crystal need be present. The
coordination of the growth rates of alternating facets is initiated by the
aerodynamics of the falling crystal.

An interesting feature of this model is that a triangular plate morphology
is both aerodynamically stable and stable against additional growth
perturbations. Once a plate takes the form of an equilateral triangle $%
(T\rightarrow 0)$, subsequent growth perturbations cannot change this
morphology, as long as the plate remains faceted. This is not true for
hexagonal plates, so even small initial perturbations would eventually
result in triangular shapes. This triangular growth instability is weak,
however, so it would likely not be a dominant effect under typical
conditions.

To understand this model in a more quantitative sense, we would have to
perform more accurate calculations of the ventilation effect for thin,
plate-like crystals. No other explanations for triangular snow crystals have
been proposed to date (to my knowledge), but the approximate analytic
methods described above are simply too crude to verify that aerodynamic
effects will indeed produce these morphologies. By combining more
sophisticated fluid mechanics calculations (e.g. see \cite{wang}) with
numerical methods for calculating crystal growth \cite{libbrechtmodel, gg,
gg3d}, it should be possible to explore this question further.

\section{Summary}

The results presented above are intended as a brief summary of aerodynamical
effects that are relevant for understanding snow crystal growth. Some
aspects of aerodynamics are quite well understood, while others are not.
Aerodynamic drag is quite well known, so calculations of terminal velocities
and related quantities are straightforward. Convection and turbulence are
more complex, however, so calculations involving these phenomena are
considerably more uncertain. 

Combining airflow and diffusion-limited growth to determine growth rates
(the ventilation effect) is difficult for complex snow crystals, in part
because their growth is limited by attachment kinetics in addition to
diffusion. Nevertheless, approximate enhancement factors are useful in
simple cases. For flow at low Reynolds number, it appears quite feasible to
combine basic computational fluid dynamics with diffusion-limited growth in
interesting cases. An especially promising route would be to use cellular
automata, as this is a very powerful technique for modeling faceted crystal
growth in 2D and 3D \cite{libbrechtmodel, gg, gg3d}. Adding low-velocity
flows to these calculations, and comparing with laboratory measurements,
would likely allow a number of interesting avenues to be explored.

The formation of triangular plate-like crystals is a particularly
interesting phenomenon to investigate further, since it appears that
aerodynamic effects are central in promoting the formation of this
morphology. Simple trigonal plates appear much more readily in some
conditions than others \cite{libbarnold}, but why this is the case is an
open question. Additional measurements at different temperatures and
supersaturations would shed light on this. More directly, \textit{in situ}
observations of electrodynamically levitated crystals \cite{electro} could
be used to examine the transition from hexagonal to triangular forms in real
time, and imposing different airflows around levitated crystals could test
many aspects of the aerodynamic model in detail.

\section{Appendix I -- Some Relevant Physical Quantities}

\noindent \textbf{Air Density.} The density of ordinary air is%
\[
\rho _{air}\approx 1.3\textrm{ }\left( \frac{P}{\textrm{1 bar}}\right) \textrm{
kg/m}^{3}
\]%
where $P$ is the air pressure. The density of ice is%
\[
\rho _{air}\approx 917\textrm{ kg/m}^{3}
\]

\noindent\textbf{Diffusion Constant.} The diffusion constant for water vapor
(or other light molecules) in air near room temperature is%
\[
D\approx 2\times 10^{-5}\textrm{ }\left( \frac{\textrm{1 bar}}{P}\right) \textrm{ m%
}^{2}\textrm{/sec} 
\]%
which increases roughly linearly with absolute temperature \cite{diffusion}.

\noindent\textbf{Air Viscosity.} The dynamic viscosity of air is%
\[
\mu \approx 1.8\times 10^{-5}\textrm{ kg/(m-sec)} 
\]%
at room temperature. The viscosity is roughly independent of pressure as
long as the molecular mean free path is short compared to other lengths in
the system. In air the molecular mean free path is roughly 0.1 $\mu $m. The
viscosity increases slowly as the temperature increases; at -10 C (263 K)
the viscosity is $\mu =1.7\times 10^{-5}$ kg/(m-sec) \cite{wikiviscosity}.

The kinematic viscosity is%
\begin{eqnarray*}
\nu _{kin} &=&\frac{\mu }{\rho _{air}} \\
&\approx &1.4\times 10^{-5}\textrm{ }\left( \frac{\textrm{1 bar}}{P}\right) 
\textrm{ m}^{2}\textrm{/sec}
\end{eqnarray*}%
From the statistical mechanics of ideal gases we have that $D\approx \nu
_{kin}$ to a rough approximation. The ratio of these quantities is known as
the Schmidt number $N_{Sc}=\nu _{kin}/D$, and experimentally in dilute gases
it has been found that $\nu _{kin}/D\approx 0.7$ \cite{reif, pruppacher}.

\section{Appendix II -- The Turbulence Spectrum}

When looking at the effects of aerodynamics on snow crystal growth, we would
like to know the motion of a crystal and the relative velocity between the
crystal and the air. One source of motion and relative velocity is gravity,
which produces the terminal velocity of a falling crystal we calculated
above. Another source of motion and relative velocity is from wind and
turbulence.

To examine the effects of turbulence, first consider a snow crystal
suspended in a parcel of air that is oscillating in time with position and
velocity given by $x_{air}=x_{0}e^{i\omega t}$ and $u_{air}=i\omega
x_{0}e^{i\omega t}.$ The drag force will cause the crystal to oscillate at
the same frequency, so we write the crystal amplitude and velocity as $%
x_{xtal}=y_{0}e^{i\omega t}$ and $u_{xtal}=i\omega y_{0}e^{i\omega t}.$ The
relative velocity of the crystal with respect to the air is then $%
u_{rel}=i\omega (x_{0}-y_{0})e^{i\omega t}.$ Using the known drag force, we
have (dropping the $e^{i\omega t})$

\begin{eqnarray*}
F_{drag} &=&bu \\
&=&i\omega b(x_{0}-y_{0})
\end{eqnarray*}%
and from $F=ma$ we have%
\[
y_{0}=\frac{i\omega b}{m\omega ^{2}+i\omega b}x_{0} 
\]%
with the relative velocity 
\[
u_{rel}=\frac{m\omega ^{2}}{m\omega ^{2}+i\omega b}u_{air} 
\]%
For $\omega \rightarrow 0$ we see $y_{0}\rightarrow x_{0}$ and $%
u_{rel}\rightarrow 0.$ In this case the crystal simply moves along with the
air. At high frequencies, the inertia of the crystal is more important and $%
y_{0}\sim x_{0}/\omega $ and $u_{rel}\rightarrow u_{air}.$ For a spherical
crystal, the transition frequency between these regimes is%
\begin{eqnarray*}
\omega _{c} &=&\frac{b}{m} \\
&\approx &\frac{9\mu }{2\rho _{ice}R^{2}}\textrm{ (spherical crystal)} \\
\nu _{c} &=&\frac{\omega _{c}}{2\pi }\approx 1.6\left( \frac{100\textrm{ }\mu 
\textrm{m}}{R}\right) ^{2}\textrm{ Hz}
\end{eqnarray*}%
For a thin disk oriented such that $R_{H}\approx R$, we have%
\begin{eqnarray*}
\omega _{c} &\approx &\frac{6\mu }{\rho _{ice}RT}\textrm{ (thin disk)} \\
\nu _{c} &=&\frac{\omega _{c}}{2\pi }\approx 20\left( \frac{100\textrm{ }\mu 
\textrm{m}}{R}\right) \left( \frac{10\textrm{ }\mu \textrm{m}}{T}\right) \textrm{ Hz}
\end{eqnarray*}%
In both cases we have $\left\vert u_{rel}\right\vert \approx \left\vert
u_{air}\right\vert $ for $\omega >\omega _{c}$, while for $\omega <\omega
_{c}$ we have 
\[
\left\vert u_{rel}\right\vert \approx \frac{\nu }{\nu _{c}}\left\vert
u_{air}\right\vert 
\]

\subsection{The Von Karman Model}

Two models are often used to describe the spectrum of atmospheric turbulence
-- the Von Karman spectrum and the Davenport spectrum \cite{turb}. Both
models display a $\nu ^{-5/3}$ Kolmogorov behavior over a broad frequency
range. The Von Karman model gives the power spectral density of the wind
velocity as \cite{turb}%
\[
S(\nu )\approx \frac{4I^{2}UL}{\left[ 1+70.8\left( \nu L/V\right) ^{2}\right]
^{5/6}} 
\]%
where $S$ is in (m/sec)$^{2}/$Hz, $\nu $ is the frequency in Hz, $U$ is the
mean wind speed in m/sec, $I$ is the turbulence intensity (between 0 and 1;
typically around 0.2), and $L$ is the outer scale of the turbulence in
meters (around 80 in the open air). Integrating this expression over
frequency gives

\begin{eqnarray*}
V_{eff}^{2} &=&\int_{0}^{\infty }S\left( \nu \right) d\nu \\
&\approx &I^{2}U^{2}
\end{eqnarray*}

For $\nu <\nu _{c}$ we can write the power spectral density for the relative
air/crystal velocity as%
\begin{eqnarray*}
S_{rel} &\approx &\left( \frac{\nu }{\nu _{c}}\right) ^{2}S \\
&\approx &\frac{4I^{2}U^{3}}{\nu _{c}^{2}C^{2}L}\frac{x^{2}}{\left(
1+x^{2}\right) ^{5/6}}
\end{eqnarray*}%
where $x=C\nu L/U$ and $C=\sqrt{70.8}.$ The low-frequency knee in $S(\nu )$
occurs at $\nu _{knee}\approx U/CL$ Hz, and taking $U\approx 1$ m/sec for a
typical (light) wind velocity gives $\nu _{knee}\approx 1$ mHz. Since $\nu
_{knee}\ll \nu _{c},$ we can assume $x\gg 1$ over frequencies we are
interested in and write%
\[
S_{rel}(\nu )\approx \frac{4I^{2}U^{3}}{\nu _{c}^{2}C^{2}L}x^{1/3} 
\]

Integrating this from $\nu =0$ to $\nu _{c}$ gives the mean-squared relative
air/crystal velocity arising from turbulence%
\begin{eqnarray*}
\left\langle v_{rel}^{2}\right\rangle &\approx &\int_{0}^{\nu
_{c}}S_{rel}\left( \nu \right) d\nu \\
&\approx &\frac{3I^{2}U^{8/3}}{C^{5/3}L^{2/3}}\nu _{c}^{-2/3} \\
&\approx &0.0002U^{8/3}\nu _{c}^{-2/3}
\end{eqnarray*}%
where $\left\langle v_{rel}^{2}\right\rangle $ is in (m/sec)$^{2}$ and we
used $I=0.2$ and $L=80$ for the last expression. Our final result for the
Von Karman turbulence model becomes%
\begin{eqnarray*}
v_{turb,RMS} &=&\left\langle v_{rel}^{2}\right\rangle ^{1/2} \\
&\approx &0.014U^{4/3}\nu _{c}^{-1/3}\textrm{ m/sec}
\end{eqnarray*}

\subsection{The Davenport Model}

For comparison, we also evaluate the Davenport model \cite{turb}, which gives%
\[
S\approx \frac{4kx^{2}U^{2}}{\nu (1+x^{2})^{4/3}} 
\]%
where the notation is similar to that used above, except now $x=1200\nu /U$
and $k$ is a roughness coefficient $(k\approx 0.08$ for open terrain).
Integrating this gives 
\begin{eqnarray*}
V_{eff}^{2} &=&\int_{0}^{\infty }S\left( \nu \right) d\nu \\
&=&6kU^{2} \\
&\approx &0.5U^{2}
\end{eqnarray*}%
The low-frequency knee occurs at $\nu _{knee}\approx V/1200$ Hz, which is
again around 1 mHz. For $\nu <\nu _{c}$ we can write%
\begin{eqnarray*}
S_{rel} &\approx &\left( \frac{\nu }{\nu _{c}}\right) ^{2}S \\
&\approx &\frac{4kU^{3}}{1200\nu _{c}^{2}}x^{1/3}
\end{eqnarray*}%
where again we have assumed $x\gg 1$ over frequencies of interest.
Integrating this as we did above gives%
\begin{eqnarray*}
\left\langle v_{rel}^{2}\right\rangle &\approx &\int_{0}^{\nu
_{c}}S_{rel}\left( \nu \right) d\nu \\
&\approx &0.002U^{8/3}\nu _{c}^{-2/3} \\
v_{turb,RMS} &\approx &0.05U^{4/3}\nu _{c}^{-1/3}\textrm{ m/sec}
\end{eqnarray*}

We see that $v_{turb,RMS}$ in the Davenport model is roughly 3 times larger
than for the Von Karman model, reflecting the fact that turbulence models
are not without considerable uncertainty. Using the expressions above for $%
\nu _{c}$, we have 
\begin{eqnarray}
v_{turb,RMS}\textrm{ (spherical crystals)} &\approx &\left( 2.7\pm 1.5\right)
\left( \frac{U}{\textrm{1 m/sec}}\right) ^{4/3}\left( \frac{R}{100\textrm{ }\mu 
\textrm{m}}\right) ^{2/3}\textrm{ cm/sec}  \label{turb1} \\
v_{turb,RMS}\textrm{ (thin plate crystals)} &\approx &\left( 1.2\pm 0.7\right)
\left( \frac{U}{\textrm{1 m/sec}}\right) ^{4/3}\left( \frac{R}{100\textrm{ }\mu 
\textrm{m}}\right) ^{1/3}\left( \frac{T}{10\textrm{ }\mu \textrm{m}}\right) ^{1/3}%
\textrm{ cm/sec}  \nonumber
\end{eqnarray}%
where again $U$ is the mean wind velocity. The uncertainty in the constants
in these expressions reflects the differences in the two turbulence models.


\begin{thebibliography}{99}
\bibitem{minerals} Prinz, M., Harlow, G., and Peters, J., \textquotedblleft
Simon \& Schuster's Guide to Rocks \& Minerals,\textquotedblright\
Publisher:Simon \& Schuster (1978).

\bibitem{nano} Imai, H., \textquotedblleft Self-organized Formation of
Hierarchical Structures,\textquotedblright\ Publisher:Springer-Verlag (2007).

\bibitem{solidification} Kolasinski, K. W., \textquotedblleft Solid
structure formation during the liquid/solid phase
transition,\textquotedblright\ Curr. Opin. Solid State \& Mater. Sci., 11,
76-85 (2007).

\bibitem{shapes} Ben-Jacob, E., and Garik, P., \textquotedblleft Ordered
shapes in nonequilibrium growth,\textquotedblright\ Physica D, 38, 16-28
(1989).

\bibitem{saito} Saito, Y., \textquotedblleft Statistical Physics of Crystal
Growth,\textquotedblright\ Publisher:World Scientific (1996).

\bibitem{fieldguide} Libbrecht, K. G., \textquotedblleft Ken Libbrecht's
Field Guide to Snowflakes,\textquotedblright\ Publisher:Voyageur Press
(2006).

\bibitem{libbrechtreview} Libbrecht, K. G., \textquotedblleft The physics of
snow crystals,\textquotedblright\ Rep. Prog. Phys., 68, 855-895 (2005).

\bibitem{libbrechtmodel} K. G. Libbrecht, \textquotedblleft Physically
derived rules for simulating faceted crystal growth using cellular
automata,\textquotedblright\ arXiv:0807.2616 (2008).

\bibitem{gg} Gravner, J. and Griffeath, D., \textquotedblleft Modeling snow
crystal growth II: A mesoscopic lattice map with plausible
dynamics,\textquotedblright\ Physica D 237, 385-404 (2008).

\bibitem{gg3d} Gravner, J. and Griffeath, D., \textquotedblleft Modeling
snow-crystal growth: A three-dimensional mesoscopic
approach,\textquotedblright\ Phys. Rev. E 79, 011601 (2009).

\bibitem{pruppacher} Pruppacher, H. R., and Klett, J. D., \textquotedblleft
Microphysics of clouds and precipitation,\textquotedblright\ Publisher:
Kluwer Academic Publishers (1997).

\bibitem{hallett} Keller, V. W. and Hallett, J., \textquotedblleft Influence
of air velocity on the habit of ice crystal growth from the
vapor,\textquotedblright\ J. Cryst. Growth 60, 91-106 (1982).

\bibitem{fukuta99} Fukuta, N. and Takahashi, T., \textquotedblleft The
growth of atmospheric ice crystals: A summary of findings in vertical
supercooled cloud tunnel studies,\textquotedblright\ J. Atmos. Sci. 56,
1963-79 (1999).

\bibitem{wang} Wang, P. K., \textquotedblleft Shape and microdynamics of ice
particles and their effects in cirrus clouds,\textquotedblright\ Adv.
Geophys. 45, 1-258 (2002).

\bibitem{wikidrag} http://en.wikipedia.org/wiki/Drag\_(physics) (2009).

\bibitem{dragcoeff} http://en.wikipedia.org/wiki/Drag\_coefficient (2009).

\bibitem{nakaya} Nakaya, U., \textquotedblleft Snow Crystals: Natural and
Artificial,\textquotedblright\ Publisher:Harvard University Press (1954).

\bibitem{naturedisks} Field, S. B., et al., \textquotedblleft Chaotic
dynamics of falling disks,\textquotedblright\ Nature 388, 252-4 (1997).

\bibitem{lightcolor} Lynch, D. K. and Livingston, W., \textquotedblleft
Light and Color in Nature,\textquotedblright\ Publisher:Cambridge University
Press (2001).

\bibitem{tape1} Tape, W. \textquotedblleft Atmospheric
Halos,\textquotedblright\ Publisher:American Geophysical Union (1994).

\bibitem{tape2} Tape, W. and Moilanen, J., \textquotedblleft Atmospheric
Halos and the Search for Angle X,\textquotedblright\ Publisher:American
Geophysical Union (2005).

\bibitem{cloudy} Libbrecht, K. G. and Tanusheva, V. M., \textquotedblleft
Cloud chambers and crystal growth:\ Effects of electrically enhanced
diffusion on dendrite formation from neutral molecules,\textquotedblright\
Phys. Rev. E 59, 3253-3261 (1999).

\bibitem{snowflakes} Libbrecht, K. G., \textquotedblleft
Snowflakes,\textquotedblright\ Publisher:Voyageur Press\ (2008).

\bibitem{libbarnold} Libbrecht, K. G. and Arnold, H. M., \textquotedblleft
Aerodynamic Stability and the Growth of Triangular Snow
Crystals,\textquotedblright\ arXiv:0911.4267 (2009). 

\bibitem{scoresby} Scoresby, W. \textquotedblleft An Account of the Arctic
Regions with a History and Description of the Northern
Whale-Fishery,\textquotedblright\ Publisher:Archibald Constable Publishing\
(1820).

\bibitem{bentley} Bentley, W. A., and Humphreys, W. J., \textquotedblleft
Snow Crystals,\textquotedblright\ Publisher:McGraw-Hill (1931).

\bibitem{yamashita} Yamashita, A., \textquotedblleft On the trigonal growth
of ice crystals,\textquotedblright\ J. Meteor. Soc. Japan 51, 307-316 (1973).

\bibitem{electro} Swanson, B. D., Bacon, M. J.; Davis, E. J., et al.,
\textquotedblleft Electrodynamic trapping and manipulation of ice
crystals,\textquotedblright\ Quart. J. Roy. Meteor. Soc., 125, 1039-1058
(1999).

\bibitem{diffusion} %
http://cambridge.org/us/engineering/author/nellisandklein/downloads/examples/EXAMPLE\_9.2-1.pdf (2009).

\bibitem{wikiviscosity} http://en.wikipedia.org/wiki/Dynamic\_viscosity
(2009).

\bibitem{turb} Bely, P.-Y., \textquotedblleft The design and construction of
large optical telescopes,\textquotedblright\ Publisher:Springer (2003).

\bibitem{reif} Reif, F., \textquotedblleft Fundamentals of statistical and
thermal physics,\textquotedblright\ Publisher:McGraw-Hill (1965).
\end{thebibliography}
\end{document}